\documentclass[sigconf]{acmart}

\usepackage{amsmath,amsfonts}
\usepackage{multirow}
\usepackage{tabularray}
\usepackage{mathrsfs}
\usepackage{makecell}
\usepackage{threeparttable}

\AtBeginDocument{%
  }

\copyrightyear{2024}
\acmYear{2024}
\setcopyright{acmlicensed}\acmConference[ICCAD '24]{IEEE/ACM International Conference on Computer-Aided Design}{October 27--31, 2024}{New York, NY, USA}
\acmBooktitle{IEEE/ACM International Conference on Computer-Aided Design (ICCAD '24), October 27--31, 2024, New York, NY, USA}
\acmDOI{10.1145/3676536.3676710}
\acmISBN{979-8-4007-1077-3/24/10}

\begin{document}

\begin{abstract}
Dynamic vision sensor (DVS) is novel neuromorphic imaging device that generates asynchronous events. Despite the high temporal resolution and high dynamic range features, DVS is faced with background noise problem. Spatiotemporal filter is an effective and hardware-friendly solution for DVS denoising but previous designs have large memory overhead or degraded performance issues. In this paper, we present a lightweight and real-time spatiotemporal denoising filter with set-associative cache-like memories, which has low space complexity of \text{O(m+n)} for DVS of $m\times n$ resolution. A two-stage pipeline for memory access with read cancellation feature is proposed to reduce power consumption. Further the bitwidth redundancy for event storage is exploited to minimize the memory footprint. We implemented our design on FPGA and experimental results show that it achieves state-of-the-art performance compared with previous spatiotemporal filters while maintaining low resource utilization and low power consumption of about 125mW to 210mW at 100MHz clock frequency.

\end{abstract}
\title{An O(m+n)-Space Spatiotemporal Denoising Filter with Cache-Like Memories for Dynamic Vision Sensors}

\author{Qinghang Zhao, Jiaqi Wang, Yixi Ji, Jinjian Wu, and Guangming Shi}

\orcid{0000-0003-0116-8975}
\affiliation{%
  \institution{Xidian University, China}
  \country{}
  \postcode{710126}
}
\email{qhzhao@xidian.edu.cn}

\renewcommand{\shortauthors}{Zhao et al.}



\maketitle
\section{Introduction}
\label{sec:intro}
Dynamic vision sensor (DVS)~\cite{4444573} is a kind of novel neuromorphic imaging device which is inspired by the principle of biological retina. The pixel of DVS asynchronously generates bipolar ON or OFF event when the change of luminous intensity of the receptive region exceeds the positive or negative threshold. Hence the output of DVS is usually represented as a quadruple ($x,y,t,p$), which are column and row coordinate of the pixel, timestamp, and polarity of the event, respectively.
From the perspective of working mechanism, the imaging process of DVS is in differential way. In contrast, conventional CMOS image sensors (CIS) are based on the integral principle, in which the capacitor in each pixel preserves the charge accumulation related to the absolute light intensity during exposure time and the charge is further converted to digital signal frame-by-frame synchronously. The period of charge accumulation constrains the frame rate to tens for typical CIS. Besides, when light intensity in some region is very high or low, the corresponding pixels in CIS are difficult to yield valid signal, resulting in a moderate dynamic range (<100dB).
In comparison, owing to its differential imaging and asynchronous readout characteristics, DVS has the advantages of high temporal resolution (\textasciitilde$\mu$s), high dynamic range (120dB - 160dB), and low power consumption~\cite{4444573,9063149}. In addition, the asynchronous output of DVS inherently fits with Spiking Neural Network (SNN). Therefore, DVS has received a lot of attention and been extensively exploited for a variety of computer vision tasks, such as object recognition~\cite{gehrig2023recurrent}, object tracking~\cite{Messikommer_2023_CVPR}, video deblurring~\cite{10471264}, video frame interpolation~\cite{9962797}, etc.

However, along with the merits, the differential imaging manner of DVS makes it more sensitive to the background activity (BA) caused by thermal noise and junction leakage currents~\cite{Guo2022}, which has significant impact on the quality of output signal and induces the communication bandwidth and power consumption overhead. Therefore, denoising is of particular significance in research of DVS.
The fundamental principle of denoising is that the valid events in the stream show spatiotemporal correlation and noise doesn't, since the motion of object is continuous in space and time while noise is random.
In consideration of this characteristic, various methods have been proposed to deal with the denoising problem of DVS.
Spatiotemporal filter is an effective way and is also easy to implement with hardware for online processing. The principal is that if the event has correlation with antecedent ones within certain space and time window, it passes the filter and is regarded as valid signal. Therefore, maintaining a record of the coordinates and timestamps of events is vital. In work~\cite{Delbruck2008}, the memory units are as as many as the sensor pixels which are used to preserve the timestamp of most-recent event for correlation assessment. This method achieves good performance but the memory overhead is significant since the space complexity is O(mn) for a DVS of $m\times n$ resolution.
The work~\cite{Guo2022} improves the spatiotemporal filter design taking into account the number of correlated events, but still is an O(mn) scheme.
In another work~\cite{Khodamoradi2021}, two memory modules are adopted. The pixels in same row share a common memory unit of one module and the incoming events belonging to this row will update the same memory unit. And it’s the same with column.
This method significantly reduces the memory complexity from O(mn) to O(m+n), but the performance is degraded.

Other denoising techniques can be categorized as offline methods. Specifically, researchers have explored solutions using probabilistic undirected graph~\cite{Wu2020}, event density~\cite{10138453}, Convolutional Neural Network~\cite{baldwin2020event}, and Graph Neural Network~\cite{9893571}. These methods require extensive computations and are very hard to implement with hardware or to run on end device in real time. We focus on the online denoising method and more detailed discussions of offline methods are beyond the scope of this paper.

In this work, we present a novel spatiotemporal filter design named \underline{C}ache \underline{L}ike \underline{F}ilter (\textbf{CLF}), to resolve the contradiction of space complexity and performance. CLF consists of symmetric Row Denoise Module and Column Denoise Module. The modules utilize set-associative cache-like memory banks. Each block of memory bank, analogous to a cache block, stores multiple events occurring in same row or column of DVS. Therefore, the number of memory block is $m+n$ for a DVS of $m\times n$ resolution. In other words, the space complexity of CLF is O(m+n). The detail of CLF design will be discussed in detail in Section~\ref{sec:design}.
The specific contributions of this work are listed as follows:
\begin{itemize}
    \item[-] We propose a novel spatiotemporal filter design, which utilizes the cache-like memories for the first time and the space complexity is O(m+n) for a DVS of $m\times n$ resolution. We prove that our design is a more generic design which includes previous spatiotemporal filter designs. Therefore, design space can be explored to investigate the optimal parameters for different scenarios. 
    \item[-] We optimize the filter design in consideration of the characteristics of DVS denoising. We design the pipeline structure for memory access which employs a read cancellation technique to reduce power consumption. Besides, based on theoretical analysis and simulations, the bitwidth of event timestamp is reduced to minimize memory footprint.
    \item[-] We implement our design on FPGA and conduct comprehensive experiment on both simulated data and recorded data of DVS. The results demonstrate that the proposed method achieves comparable and even better performance than O(mn) design while the resource utilization and power consumption remain quite low. 
\end{itemize}
The rest of this paper is organized as follows. Section~\ref{sec:background} introduces the working principle of DVS and how the spatiotemporal filters are used for denoising. Section~\ref{sec:design} presents the cache-like spatiotemporal filter design and its optimizations. Section~\ref{sec:exp} reports the experimental results. Finally, Section~\ref{sec:conclusion} concludes this paper.
\section{Background}
\label{sec:background}

\subsection{Working Principle of DVS}
\label{sec:DVSwork}
\begin{figure}[thpb]
    \centering
    \includegraphics[width=\columnwidth]{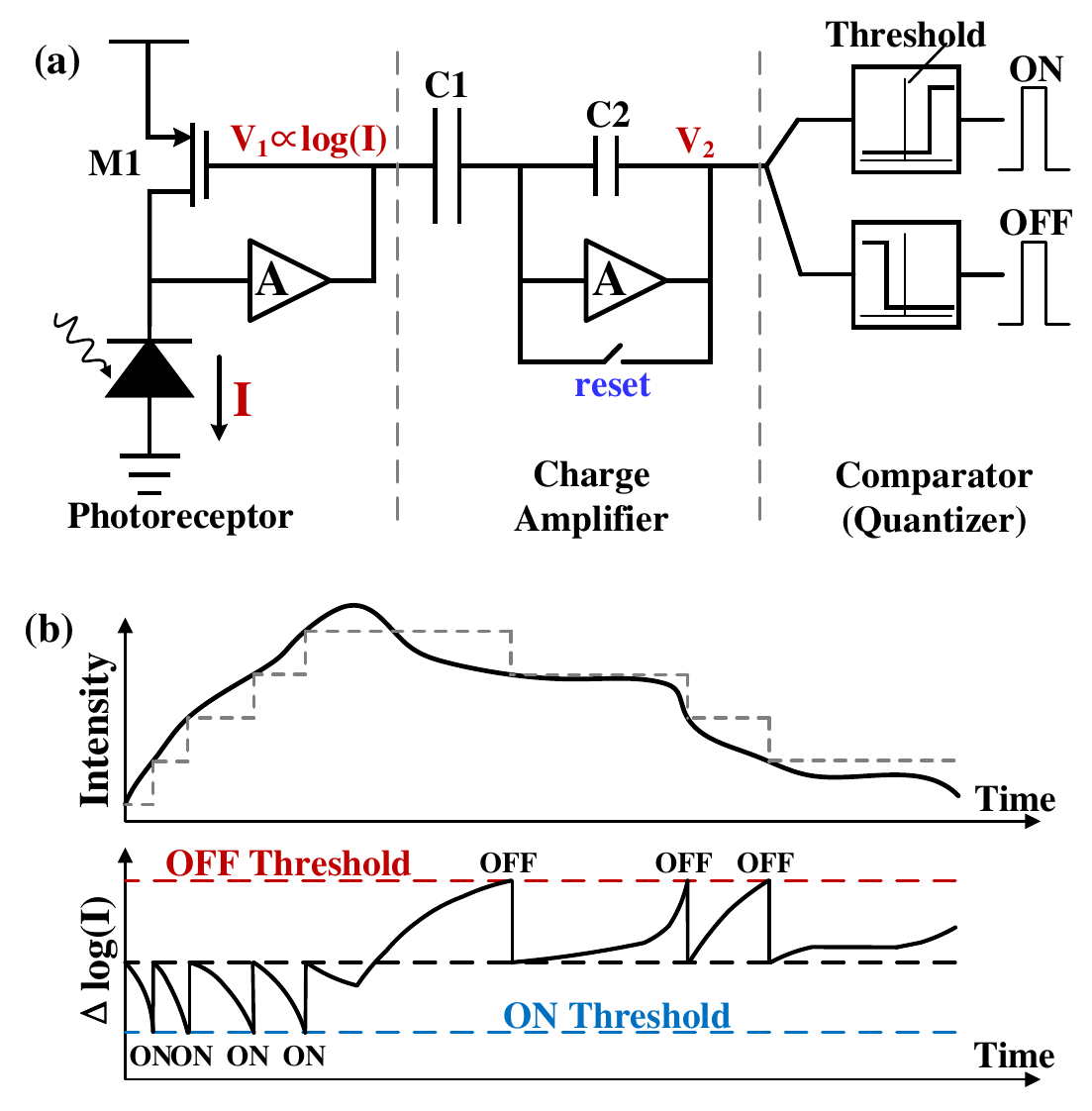}
    \vspace{-15pt}
    \caption{The (a) simplified circuit schematic and (b) operational diagram of a DVS pixel.}
    \label{fig:dvs_pixel}
\end{figure}

The simplified circuit schematic of DVS pixel is shown in Fig.~\ref{fig:dvs_pixel}(a). The pixel can be divided into three stages, which are the photoreceptor, the charge amplifier, and the comparator. The photoreceptor realizes photoelectric conversion. The MOSFET M1 in series with photodiode operates in sub-threshold region. In this way, the output voltage $V_1$ of this stage is logarithmic with respect to the photocurrent $I$, enhancing the dynamic range of DVS. The charge amplifier employs two capacitors to amplify the change of $V_1$ and its output $V_2$ is proportional to $\frac{C_1}{C_2}\Delta\log(I)$. When the output voltage of charge amplifier exceeds the upper or lower threshold, one comparator in third stage will yield high-level output, referred to as ON or OFF event. If the control logic of DVS receives the event, a pulse signal will be generated to reset the charge amplifier and its output returns to default value. The process is repeated and discrete ON/OFF event stream is generated. Fig.~\ref{fig:dvs_pixel}(b) illustrates the generation of events with the light intensity changing. As can be seen, different with the conventional CIS with fixed frame rate, DVS asynchronously generates bipolar event signal and its firing rate reflects the variation of light intensity.
It can be known through the above analysis that the differential and asynchronous nature of DVS imaging makes it high dynamic range, high time resolution, frame free, and ultra fast. Besides, since the ADC is substituted by comparator, the power consumption of DVS is significantly reduced.
\vspace{-5pt}

\vspace{-10pt}
\subsection{Spatiotemporal Filters}
\label{sec:filters}
\begin{figure}[htbp]
    \centering
    \includegraphics[width=\columnwidth]{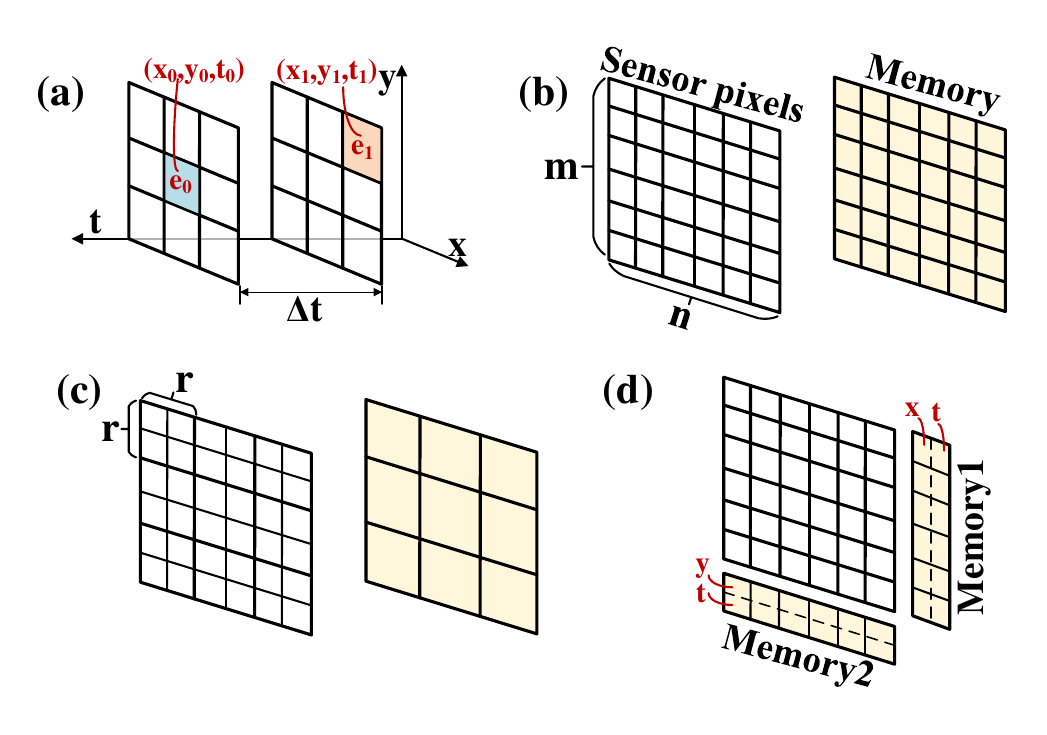}
    \vspace{-30pt}
    \caption{The (a) principle of spatiotemporal filter and (b-d) illustration of different architectures.}
    \label{fig:spatemprol_filter}
    \vspace{-10pt}
\end{figure}

Spatiotemporal filter is a method for DVS denoising, which is based on the principle that imaging of objects is highly correlated in space and time, while noise is random. As shown in Fig.~\ref{fig:spatemprol_filter}(a), it is assumed that one event $e_0$ is occurring at the current time $t_0$ at pixel coordinate ($x_0$, $y_0$) with polarity $p_0$, which is denoted by quadruple ($x_0$, $y_0$, $t_0$,$p_0$). Event $e_0$ is considered correlated with a preceding event $e_1$ denoted by ($x_1$, $y_1$, $t_1$, $p_1$), if their temporal and spatial differences fall within predefined thresholds, i.e., $t_0-t_1\leq T_{th}$, $x_0-x_1\leq D_{th}$, and 
$y_0-y_1\leq D_{th}$. The spatiotemporal filter determines whether an event is valid signal or noise according to the number of correlated events $n_{e0}$ and a preset correlation criterion, represented by the parameter $N_{cr}$. If $n_{e0}\geq N_{cr}$, the event is regarded as valid signal; otherwise, it is categorized as noise.

Spatiotemporal filters is not only straightforward in principle but also easy to implement with hardware. Various spatiotemporal filters have been designed in previous work. The elementary version is shown in Fig.~\ref{fig:spatemprol_filter}(b), in which each sensor pixel corresponds to one memory unit to store the timestamp of event. 
Delbruck proposed the background activity filter (\textbf{BAF})~\cite{Delbruck2008}. The timestamp of incoming event $e_0$ compares to that of the corresponding memory unit to determine whether it is noise or not. And then it will update the timestamp of $3\times3$ nearest-neighbor memory units. Afterwards, Linares-Barranco et al.~\cite{Linares-Barranco2015,8836544} implemented BAF on FPGA.
It can be seen that in BAF, both the spatial threshold $D_{th}$ and the correlation criterion $N_{cr}$ is 1.
Guo et al.~\cite{Guo2022} proposed Spatiotemporal Correlation Filter (STCF), in which $N_{cr}$ can vary from 1 to 8 for $3\times 3$ neighborhood, making the design more adaptive. 
However, both BAF and STCF are O(mn) designs in terms of space complexity.
When sensor resolution gets high, the memory overhead gets larger.
For example, with 4 byte timestamp, the memory size for DVS of 1920$\times$1080 resolution ~\cite{chen2019live} will be about 3.9MB, which is rather expensive to implement on chip.
An improved spatiotemporal filter design is depicted in Fig.~\ref{fig:spatemprol_filter}(c). This scheme is referred to as Subsampling Shared Memory (SSM) because $r\times r$ sensor pixels share one memory unit using subsampling technique. Except the memory organization, SSM works in similar way to BAF and STCF. SSM reduces the memory footprint by a factor of $r^2$, but still requires O(mn) memory. Khodamoradi et al.~\cite{Khodamoradi2021} proposed a spatiotemporal filter with O(n) space complexity for the first time, which is known as the first noise filter that scales less than O(n$^2$) in memory cost.
As illustrated in Fig.~\ref{fig:spatemprol_filter}(d), two memory modules are adopted in this scheme.
Each row of pixels share one memory unit of Memory1, which stores column coordinate ($x$) and timestamp ($t$) of incoming event in that row. Memory2 works in similar way.
Considering the memory architecture, we call it Row and Column Filter (\textbf{RCF}) for simplicity.
The researchers also realized RCF on FPGA and showed that is was effective for sparse DVS stream.
The RCF scheme greatly reduces memory requirement.
However, when imaging scene gets complex in which multiple objects or multiple parts of object are in motion, RCF is not competent in correctly distinguishing signal and noise because only one memory unit for each row or column is insufficient to record the past events.
In summary, previous spatiotemporal filter designs have made some valuable explorations for DVS denoising. However, the design space has not been thoroughly explored, and the trade-off between denoising performance and memory overhead requires further considerations. In Section~\ref{sec:design}, we will elaborate on our design to address the above research questions.
\section{Cache-Like Spatiotemporal Filter Design}
\label{sec:design}

\subsection{Overall Architecture}
\label{sec:filter_deisgn}

\begin{figure*}[thpb]
    \centering
    \includegraphics[width=\textwidth]{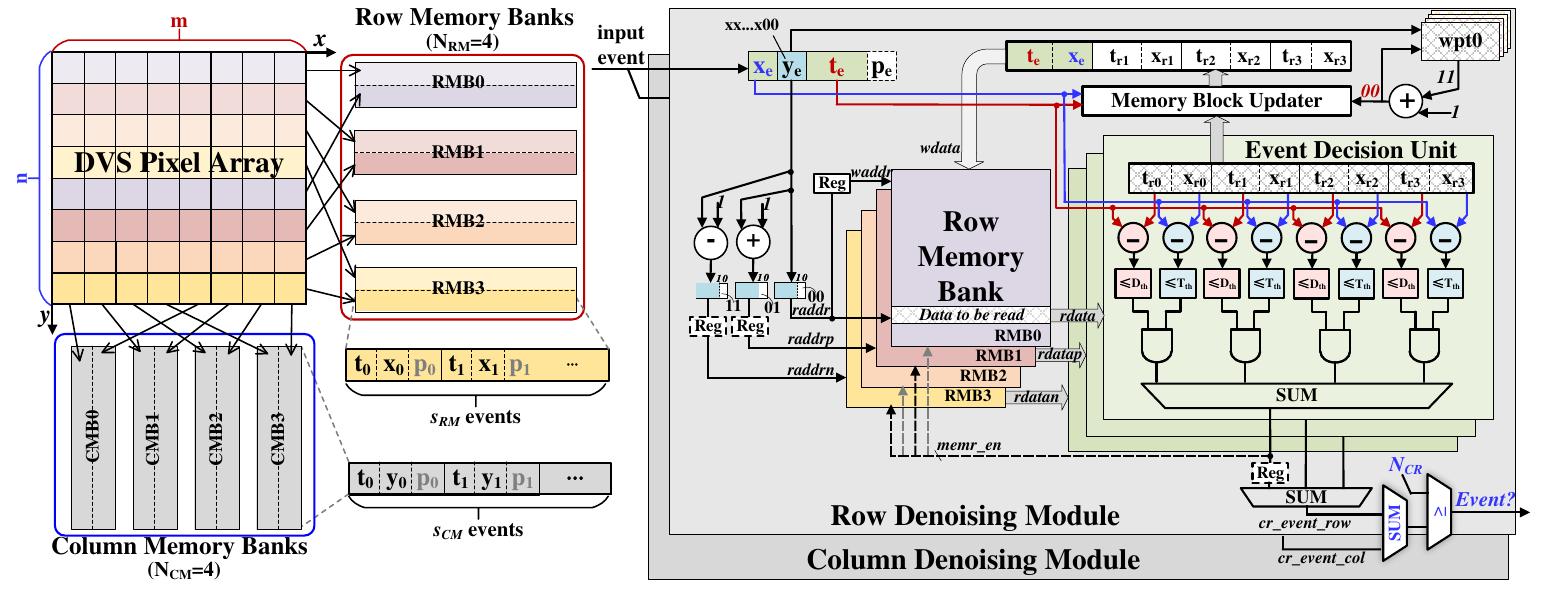} 
    \vspace{-20pt}
    \caption{The overall architecture of the proposed spatiotemporal denoising filter.}
    \label{fig:arch_filter}
\end{figure*}

\begin{table}[h]
\renewcommand\arraystretch{1.2}
\centering
\caption{The List of Symbols and Notations.}
\label{tab:symbols}
\vspace{-10pt}
\begin{tabular}{{c}p{6.5cm}}
\hline
\textbf{Symbol} & \textbf{Description}                                                            \\ \hline
$N_{RM}$    & Number of Row Memory Banks                                             \\ \hline
$N_{CM}$    & Number of Column Memory Banks                                          \\ \hline
$s_{RM}$    & \makecell[l]{Number of events that can be stored in each block \\ of row memory bank}    \\ \hline
$s_{CM}$    & \makecell[l]{Number of events that can be stored in each block \\ of column memory bank} \\ \hline
$D_{th}$    &{Threshold to decide whether events are spatially correlated}            \\ \hline
$T_{th}$    & \makecell[l]{Threshold to decide whether events are temporally \\ correlated}           \\ \hline
$N_{CR}$    & \makecell[l]{Threshold of the number of correlated events to \\ decide event or noise}  \\ \hline
$BW_T$    & Bitwidth of timestamp of event                                         \\ \hline
\end{tabular}
\end{table}

Fig.~\ref{fig:arch_filter} shows the overall architecture of the proposed lightweight and hardware-friendly spatiotemporal denoising filter, referred to as CLF.
The symbols used in this paper are listed in Table~\ref{tab:symbols}.
The CLF consists of two symmetric modules: the Row Denoising Module (RDM) and the Column Denoising Module (CDM), which work in similar way. In each module, the memory organization is cache-like and that's the meaning of CLF.
We take RDM as example to explain our design in detail.
As Fig.~\ref{fig:arch_filter} shows, assuming the resolution of DVS is $m\times n$, RDM includes $N_{RM}$ independent memory banks, designated as RMB0, RMB1, $\dots$, RMB[$N_{RM}$-1]. Each bank consists of $ \left \lceil  \frac{n}{N_{RM}} \right \rceil $ memory blocks, which are analogous to cache blocks. 
The memory is orchestrated in analogous to a set-associative cache.
If the row coordinate of one event is $y_e$, the index of Row Memory Bank to be stored is $y_e\mod N_{RM}$. 
For example, if $N_{RM}=4$, the last two bits of coordinate $y$ is exact the index of the memory set.
As Fig.~\ref{fig:arch_filter} shows, the events generated from pixels in row 0, 4, $\dots$, will be stored in RMB0, events in row 1, 5, $\dots$, in RMB1, and so on. 
And the index of memory block to store the event is $\left \lfloor \frac{y}{N_{RM}}  \right \rfloor $.
Each memory block stores $s_{RM}$ events arising from corresponding row.
In our design, the storage location of each event in memory block is not constrained. In this way, each Row Memory Bank behaves in similar way with an $s_{RM}$-way set-associative cache. The difference is that in traditional set-associative caches, each set consists of multiple cache blocks while in the Row Memory Bank, each memory block stores multiple events.
Since the mapping relation between memory bank and memory block and pixels from certain row is explicit, the row coordinate $y_e$ does not need to be saved. Therefore, the timestamp $t_e$, column coordinate $x_e$, and polarity $p_e$, if necessary, are to be stored in memory block. Usually $s_{RM}$ is much less than $n$. Therefore, replacement policy of event in memory block should be considered. Our design employs FIFO-based replacement strategy. First, it is hardware-friendly. Besides, since the timestamp of current event is no less than that of previous ones and more recent events are more likely to be used for correlation judgement in future due to spatial and temporal locality, FIFO policy behaves very much like LRU for DVS.
To indicate which previous event will be replaced, an index pointer for each memory block is required, of which the bitwidth is $\left \lceil log_2 s_{RM} \right \rceil$. In our design, a small memory module named \textbf{wpt} is adopted for each row memory bank, which contains the index pointer for each memory block.
Column memories are organized in similar way with row memories and the timestamp and column index ($x_e$) of the current event are stored in memory block. We will leave the details to avoid repetition. 

The right part of Fig.~\ref{fig:arch_filter} illustrates the architecture of CLF and the workflow for denoising. In this exemplary design, four Row Memory Banks are utilized ($N_{RM}=4$) and four events can be stored in each memory block ($s_{RM}=4$). $D_{th}$ is set to be 1, which means the previous events in $3\times 3$ spatial window should be investigated. More specifically, the adjacent two rows and two columns are checked.
Assume the current input event is represented as ($x_e, y_e, t_e, p_e$) and lats two bits of $y_e$ is 00 in binary. $x_e$ requires $BW_x=\left\lceil log_2n\right \rceil$ bits for storage, and $y_e$ needs $BW_y=\left\lceil log_2m\right \rceil$ bits. First, $y_e-1$ and $y_e+1$ are calculated, of which the last two bits are 11 and 01, respectively. Therefore, RMB0, RMB3, and RMB1 are to be read and the read address is most significant $BW_y-2$ bits of $y_e$, $y_e-1$ and $y_e+1$, respectively. The read data from RMB0, \textit{rdata}, contains information of four previous events.
This rdata is then exported to Event Decision Unit (EDU), which first decides whether the input event is correlated with stored events by comparing the column coordinate and timestamp difference and then sums the correlated ones. 
In the meantime, the input event replaces previous one from the memory unit in RMB0 on basis of FIFO policy. The replacement position is indicated by wpt0 which increments by 1 at each time of memory update.
For example, if wpt0 is (11)$_2$, it means that last update position is 3 and therefore the input event will be written to position 0. 
Two other two EDUs read data (\textit{rdatap} and \textit{rdatan}) from RMB3 and RMB1 and determine whether adjacent rows of input event are carrying correlated events and calculate the sums. RDM combines the output of three EDUs to generate total number of correlated events in all three memory blocks. CDM works in the same way as RDM. 
The output of RDM and CDM are summed up and then compared to the threshold parameter $N_{CR}$ to generate final result of CLF.
It can be known that the proposed spatiotemporal filter can process an input event and yield result in four clock cycles, indicating that our design is a real-time scheme.

From the above analysis, it can be known that for $3\times 3$ spatial window ($D_{th}=1$), three memory blocks should be simultaneously accessed for both row and column memories. To realize this requirement, at least three set of memory banks should be used. We employ four memory banks rather than three for that these banks can be direct distinguished by the last two bits of reference address and the modular operation of 3 is avoided. Similarly, if $s\times s$ spatial window is adopted, the number of memory banks should be set to $2^{\left \lceil log_2s  \right \rceil }$.

In fact, different spatiotemporal filters introduced in Section~\ref{sec:filters} can be regarded as specific design cases of CLF. If $N_{RM}$, $N_{CM}$, $s_{RM}$, and $s_{CM}$ are all set to 1, CLF degrades into RCF. 
When CDM is not used, $N_{RM}$ is set to 1, and $s_{RM}$ is set to $m$, but the stored location of event is constrained to specific position of each memory block, CLF degrades into BAF or STCF. Further, if multiple rows of pixels are mapped to the same memory block, CLF becomes SSM. 
From this perspective, our spatiotemporal filter design is more generic and therefore design space exploration will be more effective.

\vspace{-5pt}
\subsection{Pipeline Architecture for Memory Access}
\label{sec:pipeling}
\begin{figure}[htbp]
    \centering
    \includegraphics[width=\columnwidth]{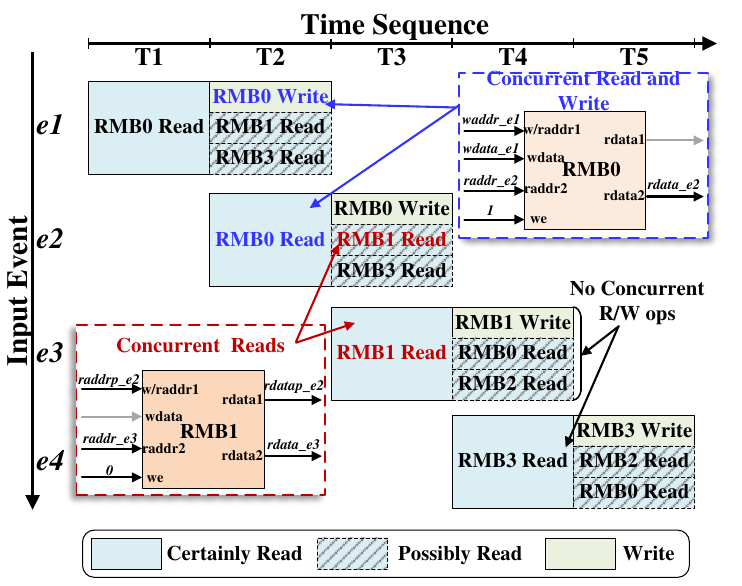}
    \vspace{-15pt}
    \caption{The two-stage pipeline of memory access.} 
    \label{fig:mempipeline}
\end{figure}

For $3\times 3$ spatial window, theoretically the correlation of the input is more likely confirmed with the events in the same row or column since they account for $5/9$ of the neighbors. Considering this, we design the two-stage pipeline architecture for memory access. 
As Fig.~\ref{fig:arch_filter} shows, RDM can optionally adopt three registers, which are marked with dashed box, for cycle delay of data and pipelining realization. 
Assume that four events, denoted as $e1$-$e4$, stream in CLF and the corresponding memory blocks reside in RMB0, RMB1, and RMB3, respectively, the memory access process is illustrated in Fig.~\ref{fig:mempipeline}. In the first stage of processing $e1$, RMB0 is read which is determined by $y_{e1}$. In the second stage of $e1$, the memory block of RMB0 read in the first stage is updated and memory blocks of RMB1 and RMB3 correspond to $y_{e1}-1$ and $y_{e1}+1$, respectively, are read. In the meantime, the first-stage read of $e2$ is proceeded, i.e. RMB0 read. Subsequent read and write operations of row memory banks involve a similar process. 
In the process, if the output of EDU generated in the first stage is not zero which means correlated event exists, the read operation of the other two memory banks are cancelled. Otherwise, the read goes on. This technique of read cancellation is quite beneficial to reducing the energy consumption, especially in ASIC implementation. 
We know from the above explanation that the pipeline structure is only applicable to small $N_{CR}$ cases because the read cancellation leads to imprecise summation result of EDU. Nevertheless, $N_{CR}=1$ is effective enough, as will be shown in Section~\ref{sec:exp}.
It should be noted that concurrent reads or concurrent read and write should be taken into consideration, as Fig.~\ref{fig:mempipeline} shows. We make use of dual-port memory to meet the requirement of concurrent access. CDM works in same way with RDM.
Despite the implementation of the pipeline architecture, the delay time of CLF is limited to only 5 clock cycles.

\vspace{-10pt}
\subsection{Bitwidth Reduction of Event Timestamp}
\label{sec:bit_reduction}
\begin{figure}[htpb]
    \centering
    \includegraphics[width=\columnwidth]{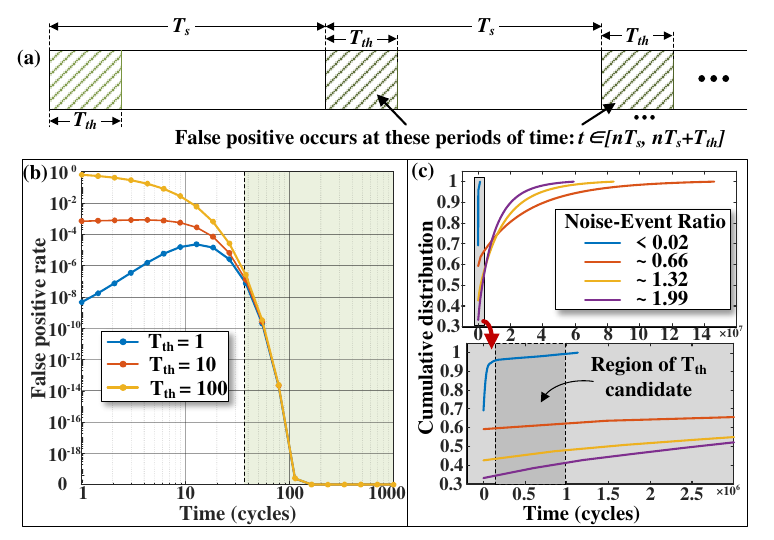}
    \vspace{-20pt}
    \caption{The relation of the bitwidth of timestamp and false positive rate, indicating the potential for bitwidth reduction.}
    \label{fig:bit_reduction}
\end{figure}
In previous work~\cite{Khodamoradi2021, Linares-Barranco2015}, the bitwidth of timestamp stored in memory remains the same as input event, which is usually 4 byte. In fact, the timestamp field occupies the primary part of memory block since either row or column coordinate of DVS can be represented within 11 bits for sensor of resolution no more than $2048\times 2048$. As shown in Fig.~\ref{fig:bit_reduction}(a), assume $T_{th}$ is the time threshold to discriminate event and noise and $T_s$ is range that the timestamp field can represent, then when the first of the subsequent input events in spatial window falls in the time range of [$n\!\cdot T_s$, $n\!\cdot T_s+T_{th}$], noise is misjudged as signal, or in other words false positive occurs. Previous works~\cite{Guo2022, Khodamoradi2021} show that the DVS noise generally complies with Poisson distribution.
Taking the Poisson probability density function into consideration, we ran simulations to inspect the relation of false positive rate and time. It should be noted that the time here is in form of clock cycles for general expression. As can be seen from Figure~\ref{fig:bit_reduction}(b), when time exceeds a certain limit, the false positive rate can be negligible for different value of $T_{th}$. This indicates that the bitwidth of $T_s$ is no need to be very large. We also collected the statistics of the time difference between input event and the most recent one in $3\times 3$ spatial window. The result in Fig.~\ref{fig:bit_reduction}(c) shows that different value of noise to event ratio has little impact on $T_{th}$. 

\section{Experimental Evaluation}
\label{sec:exp}

\subsection{Validation with Simulated Data}
\label{sec:exp_sim}

\begin{figure*}[htpb]
    \centering
    \includegraphics[width=0.9\textwidth]{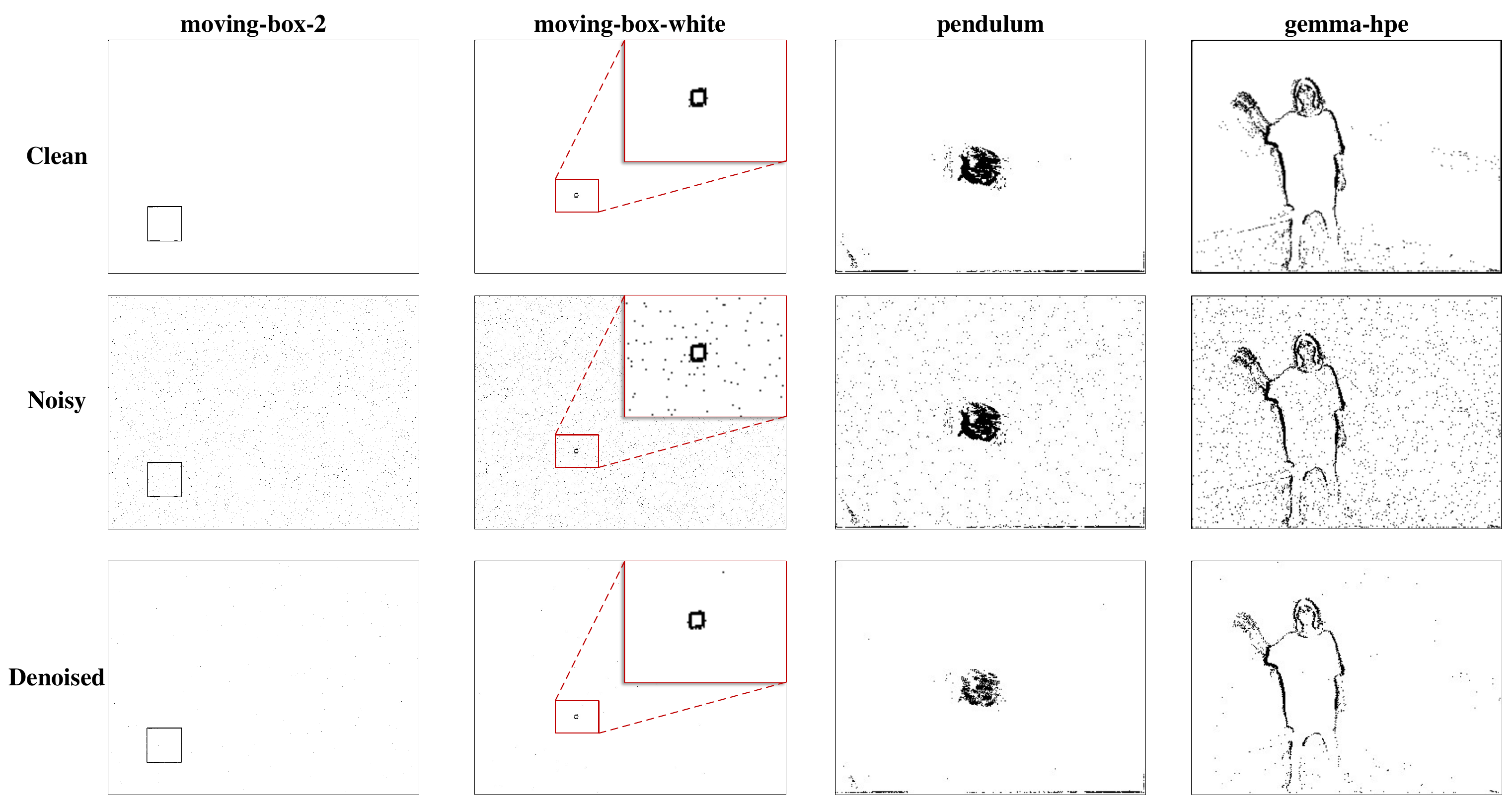}
    \vspace{-10pt}
    \caption{Visualization of clean and noisy event data generated with v2e simulator and denoised data with CLF.}
    \label{fig:sim_data}
\end{figure*}

\begin{figure}[htpb]
    \centering
    \includegraphics[width=\columnwidth]{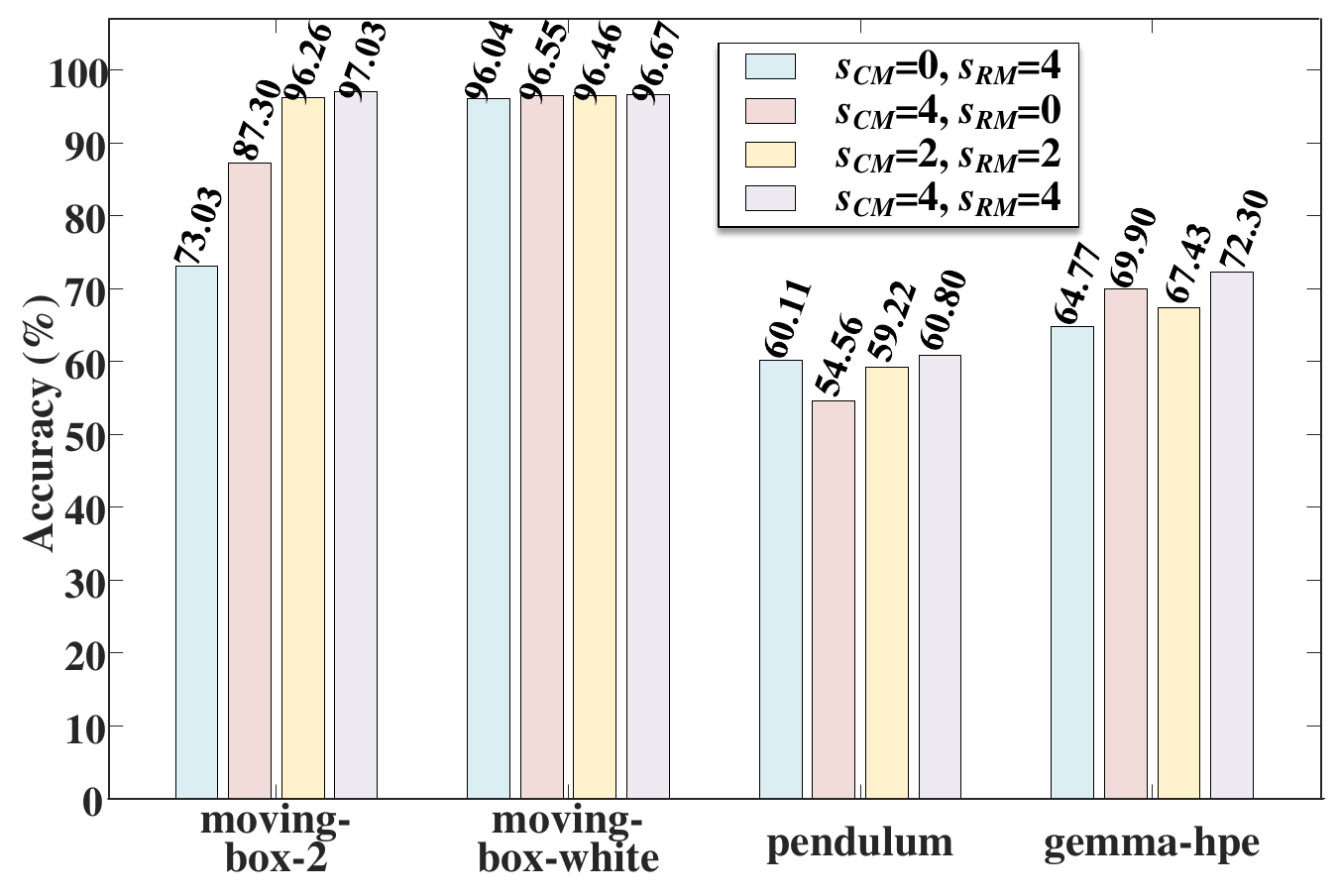}
    \caption{The denoising results of CLF in relation to the number of events that can be stored in row memory block and column memory block ($D_{th}$=1, $T_{th}$=200$\mu$s, $N_{CR}$=1).}
    \label{fig:plot_sets}
\end{figure}

\begin{figure}[htpb]
    \centering
    \includegraphics[width=\columnwidth]{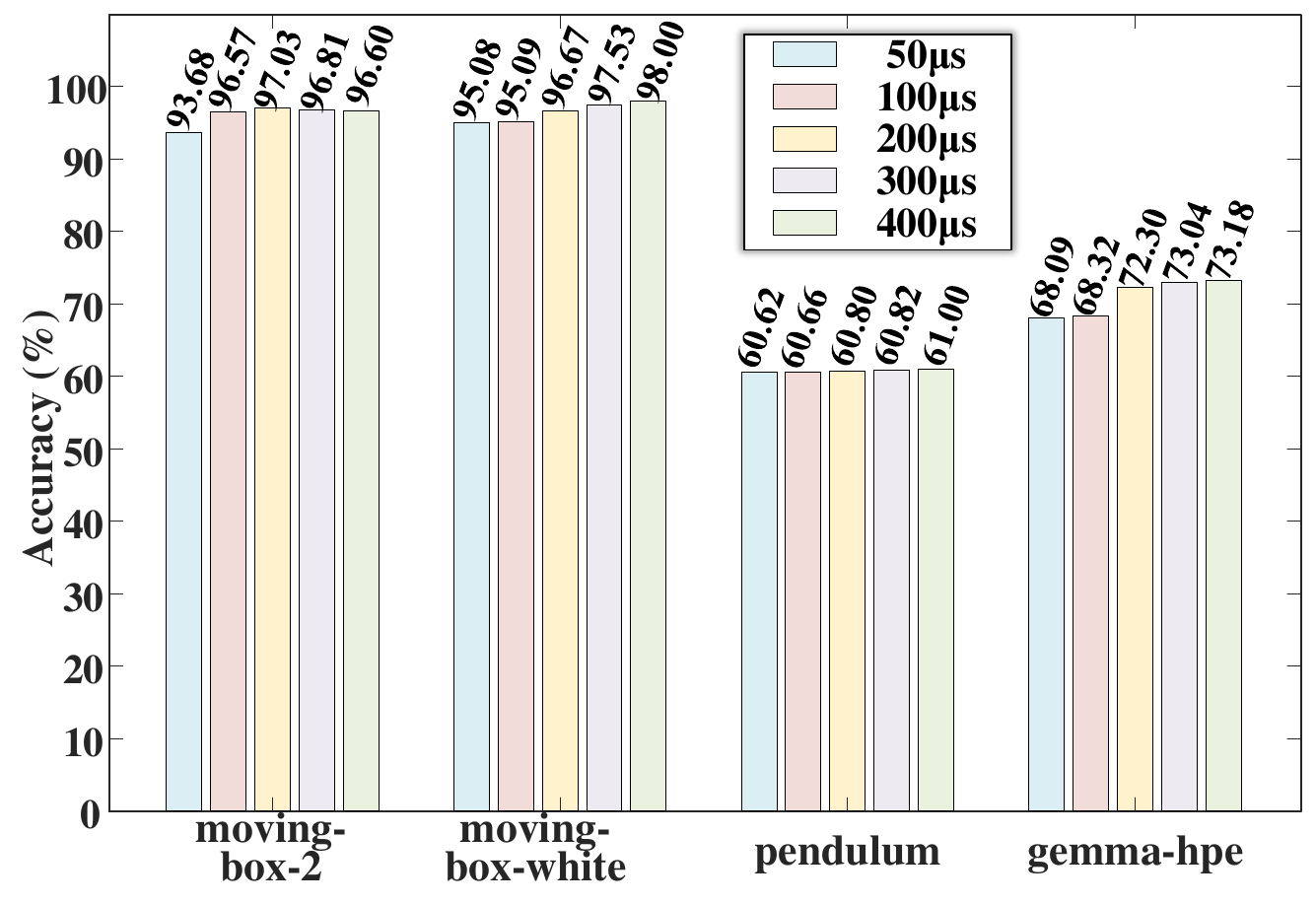}
    \caption{The denoising results of CLF in relation to the time threshold $T_{th}$ ($D_{th}$=1, $N_{CR}$=1, $s_{RM}$=4, $s_{CM}$=4).}
    \label{fig:plot_time}
\end{figure}

In this section, we will validate the effectiveness of our spatiotemporal filter design. 
We implemented CLF on Xilinx's (now AMD) low-cost XC7A35T FPGA of Artix-7 family at 100MHz clock, which means the delay time is only 50ns.
We realized various versions to cover different design configurations, including $N_{RM}$/$N_{CM}$, $s_{RM}$, $s_{CM}$, and $BW_{T}$. 
All implementations support DVS of maximum $1280\times 800$ resolution. 
The resource utilization and power consumption of  different design configurations are listed in Table~\ref{tab:implementation}. 
Among all the implementations, the configuration of $N_{RM}$/$N_{CM}$=4, $s_{RM}$=4, $s_{CM}$=4, and $BW_{T}$=32 requires the most resources. However, it is still quite moderate in number. 
Besides, if 8-bit timestamp is used, more than 40\% resources are saved compared with 32-bit version.
On the whole, our design is resource efficient, making it applicable on low-end FPGA devices or to implement with ASIC.
The total power consumption for different configurations ranges from 125mW to 210mW, which is quite low for FPGA.

\begin{table}[htbp]
\renewcommand\arraystretch{1.2}
\centering
\caption{Resource Utilization and Power Consumption of FPGA Implementation of CLF.}
\label{tab:implementation}
\begin{threeparttable} 
\begin{tabular}{ccccc}
\hline
\multirow{2}{*}{Configuration\tnote{1}} & \multicolumn{3}{c}{Resource Utilization} & \multirow{2}{*}{\begin{tabular}[c]{@{}c@{}}Power\tnote{2}\\ (mW)\end{tabular}} \\ \cline{2-4}
                               & LUT     & LUTRAM   & FF      &                                                                       \\ \hline
4-4-0-32                             & 5073    & 2624     & 2009    & 125                                                                   \\ \hline
4-0-4-32                             & 5681    & 2560     & 2502    & 131                                                                   \\ \hline
4-2-2-32                          & 6506    & 2592     & 3156    & 138                                                                   \\ \hline
4-4-4-32                         & 10756   & 5184     & 4442    & 180                                                                   \\ \hline
4-4-4-8                     & 5894    & 2112     & 2880    & 131                                                                   \\ \hline
8-4-4-32                      & 8205    & 2972     & 3503    & 210                                                                   \\ \hline
8-4-4-8                & 4990    & 1244     & 2500    & 140                                                                   \\ \hline
\hline
Available                      & 20800   & 9600     & 41600   & \textbackslash{}                                                      \\ \hline
\end{tabular}
 \begin{tablenotes}
        \footnotesize    
        \item[1] The configuration is in the format of $N_{RM}$/$N_{CM}$-$s_{RM}$-$s_{CM}$-$BW_{T}$. $s_{CM}=0$ means only RDM is utilized and $s_{RM}=0$ CDM.
        \item[2] Estimated with Vivado based on the implemented design.
      \end{tablenotes}
    \end{threeparttable} 
\end{table}

Due to the characteristics of DVS, the number of events output by DVS is much larger than image frames generated with traditional CIS for same scene and recording time. For example, tens of million events will be generated in several seconds with CeleX-V in typical scenarios. Therefore, it is almost impossible to label the event data to provide ground truth.
In consideration of this challenge, we first conduct experiments with simulated event data, which can be well controlled to be clean or noisy. 
We employ the \textbf{{v2e}} simulator~\cite{hu2021v2e}, which synthesizes realistic DVS data from conventional frame based video using an accurate DVS pixel model that includes DVS nonidealities. We select four videos which are also provided by v2e, including box-moving-2, box-moving-white, pendulum, and gemma-hpe, covering low to high complexity of scene. The resolution of the former two videos is $800\times 600$ and the latter two is $346\times 260$. We alter the simulator parameters to obtain the clean and noisy data with different noise-to-signal ratio. The visualization of clean and noisy event data generated with v2e simulator is shown in Fig.~\ref{fig:sim_data}. 

\begin{figure}[t]
    \centering
    \includegraphics[width=\columnwidth]{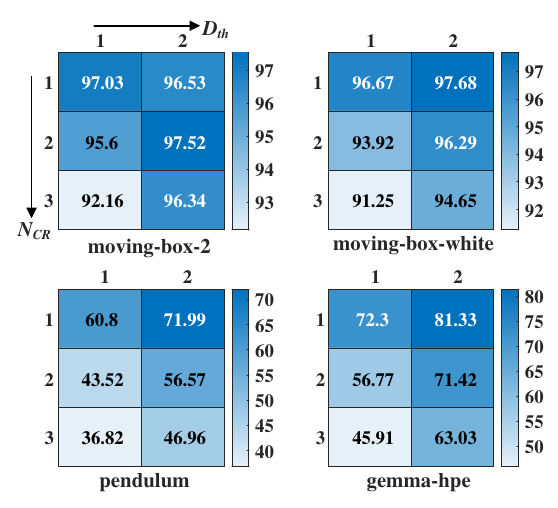}
    \vspace{-20pt}
    \caption{The denoising results of CLF in relation to $N_{CR}$ and $D_{th}$ ($s_{RM}$=4, $s_{CM}$=4, $T_{th}$=200$\mu$s).}
    \label{fig:plot_ndncr}
\end{figure}

We altered the design parameters of CLF and carried out thorough denoising experiments. Here, DVS denoising is regarded as classification problem and precision (\textbf{P}), recall (\textbf{R}), and accuracy (\textbf{A}) are used for evaluation metrics. 
To investigate the influence of parameters more apparently, we present typical results in graphs.
Fig.~\ref{fig:plot_sets} depicts the denoising results under different value of $s_{CM}$ and $s_{RM}$, i.e., whether RDM and CDM are both used and how many events can be stored in each memory block. We can see that if either RDM or CDM is used, it is uncertain which one is more effective. In other words, the event denoising problem is directional. We think it is due to the directionality of moving object's trajectory. 
We can also observe that under certain degree of resource constrains ($s_{CM}$+$s_{RM}$=4), utilizing both RDM and CDM ($s_{CM}$=2 and $s_{RM}$=2) may not ensure the optimal results (e.g. gemma-hpe), but achieves balanced performance for diverse scenarios.
Increasing the capacity of memory block ($s_{CM}$=4 and $s_{RM}$=4) can improve the performance of CLF, even though the marginal effect is diminishing. 
Fig.~\ref{fig:plot_time} shows the influence of threshold time to decide event correlation. We can see that as $T_{th}$ increases, the performance is improved slowly and approaches to a flat and even slightly degrades. For the adopted datasets, $T_{th}$ of 200$\mu$s to 400$\mu$s is sufficient.
The impact of the $D_{th}$ and $N_{CR}$ is shown in Fig.~\ref{fig:plot_ndncr}. Generally, spatial window of $5\times 5$ ($D_{th}=2$) achieves higher accuracy than $3\times 3$ ($D_{th}=1$), which is consistent with intuition. On the contrary, increasing $N_{CR}$ usually induces accuracy decline. However, it doesn't mean that larger $N_{CR}$ is of no use at all. In fact, more strict criterion is less likely to mistake noise as valid signal. From another point of view, $N_{CR}$=1 is efficient for typical scenarios.

\begin{figure*}[thpb]
    \centering
    \includegraphics[width=\textwidth]{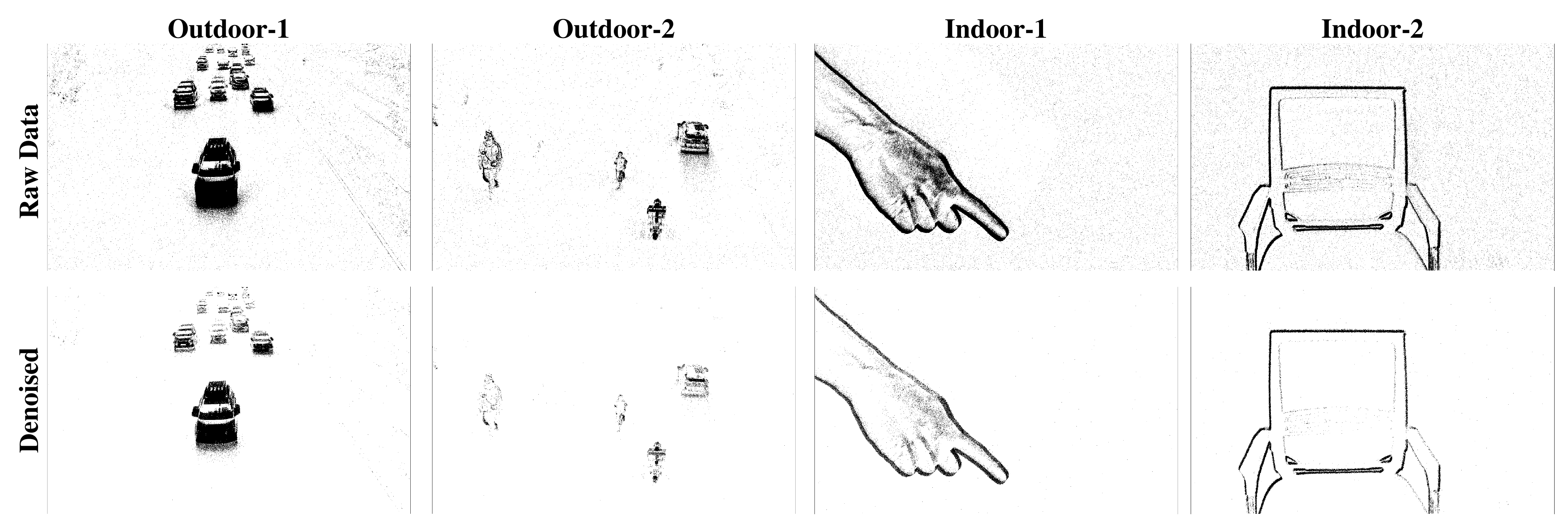}
    \vspace{-15pt}
    \caption{Visualization of recorded raw DVS data and denoising results of CLF.}
    \label{fig:real_data}
\end{figure*}

\begin{table*}[th]
\renewcommand\arraystretch{1.1}
    \caption{The Denoising Results of Different Spatiotemporal Filters.}
    \label{tab:sim_comps}
        \centering
    \begin{tblr}{
  cells = {c},
  cell{1}{1} = {c=2,r=3}{},
  cell{1}{3} = {c=6}{},
  cell{1}{9} = {c=6}{},
  cell{2}{3} = {c=3}{},
  cell{2}{6} = {c=3}{},
  cell{2}{9} = {c=3}{},
  cell{2}{12} = {c=3}{},
  cell{4}{1} = {c=2}{},
  cell{5}{1} = {c=2}{},
  cell{6}{1} = {r=4}{},
  cell{10}{1} = {c=2,r=3}{},
  cell{10}{3} = {c=6}{},
  cell{10}{9} = {c=6}{},
  cell{11}{3} = {c=3}{},
  cell{11}{6} = {c=3}{},
  cell{11}{9} = {c=3}{},
  cell{11}{12} = {c=3}{},
  cell{13}{1} = {c=2}{},
  cell{14}{1} = {c=2}{},
  cell{15}{1} = {r=4}{},
  vlines,
  hline{1,4-6,10,13-15,19} = {-}{},
  hline{2-3,11-12} = {3-14}{},
  hline{7-9,16-18} = {2-14}{},
}
\textbf{Method}                 &                    & \textbf{moving-box-2} &       &                &                      &       &                & \textbf{moving-box-while} &       &                &                       &       &                \\
                                &                    & \#Noise/\#Signal=1.29  &       &                & \#Noise/\#Signal=6.44 &       &                & \#Noise/\#Signal=5.47      &       &                & \#Noise/\#Signal=16.44 &       &                \\
                                &                    & P(\%)                 & R(\%) & A(\%)          & P(\%)                & R(\%) & A(\%)          & P(\%)                     & R(\%) & A(\%)          & P(\%)                 & R(\%) & A(\%)          \\
\textbf{BAF~\cite{Delbruck2008,Linares-Barranco2015}}                    &                    & 97.93                 & 95.31 & \textbf{97.07} & 87.39                & 86.27 & \textbf{96.48} & 97.37                     & 91.8  & \textbf{98.35}          & 80.35                 & 79.28 & \textbf{98.59} \\
\textbf{RCF~\cite{Khodamoradi2021}}                    &                    & 97.71                 & 79.53 & 90.24          & 82.76                & 67.12 & 93.7           & 96.09                     & 79.68 & 87.12          & 70.44                 & 67.46 & 97.85          \\
{\textbf{CLF}} & \textbf{$BW_T$=32, $s_{RM}/s_{CM}$=2} & 98.06                 & 93.30 & 96.26          & 83.87                & 84.92 & 95.78          & 98.21                     & 87.17 & 97.77          & 89.18                 & 82.95 & 98.44          \\
                                & \textbf{$BW_T$=32, $s_{RM}/s_{CM}$=4} & 97.97                 & 95.17 & 97.03          & 87.47                & 85.76 & 96.44          & 98.23                    & 88.62 & {98.00} & 89.32                 & 85.11 & 98.56 \\
                                & \textbf{$BW_T$=8, $s_{RM}/s_{CM}$=2}  & 98.47                 & 86.81 & 93.64          & 87.96                & 78.51 & 95.67          & 98.16                     & 84.89 & 97.42          & 88.09                & 83.81 & 98.42          \\
                                & \textbf{$BW_T$=8, $s_{RM}/s_{CM}$=4}  & 98.2                  & 88.97 & 94.46          & 84.59                & 86.21  & 96.04          & 97.42                     & 87.63 & 97.73          & 87.42                 & 87.42 & 98.56          \\
\textbf{Method}                 &                    & \textbf{pendulum}     &       &                &                      &       &                & \textbf{gemma-hpe}        &       &                &                       &       &                \\
                                &                    & \#Noise/\#Signal=0.51  &       &                & \#Noise/\#Signal=1.69 &       &                & \#Noise/\#Signal=0.41      &       &                & \#Noise/\#Signal=1.63  &       &                \\
                                &                    & P(\%)                 & R(\%) & A(\%)          & P(\%)                & R(\%) & A(\%)          & P(\%)                     & R(\%) & A(\%)          & P(\%)                 & R(\%) & A(\%)          \\
\textbf{BAF~\cite{Delbruck2008,Linares-Barranco2015}}                    &                    & 98.54                 & 40.92 & 60.39          & 90.62                & 41.17 & 76.55          & 82.58                     & 25.4  & 69.56          & 68.73                 & 33.51 & 68.89          \\
\textbf{RCF~\cite{Khodamoradi2021}}                    &                    & 98.36                 & 34.92 & 56.43          & 90.14                & 34.88 & 74.38          & 97.79                     & 46.17 & 60.99          & 84.1                  & 45.01 & 75.83          \\
{\textbf{CLF}} & \textbf{$BW_T$=32, $s_{RM}/s_{CM}$=2} &  99.04 &39.21 &59.39          & 95.37& 38.15& 76.31          & 97.98& 56.55& 68.27          & 90.01 &52.62& 79.74          \\
                                & \textbf{$BW_T$=32, $s_{RM}/s_{CM}$=4} &99.05 &41.66& 61.00          & 95.60& 41.13& 77.40 & 97.94& 63.62& \textbf{73.18}          &90.01 &61.30 &\textbf{82.68}         \\
                                & \textbf{$BW_T$=8, $s_{RM}/s_{CM}$=2}  & 98.87 &43.04 &61.85          & 92.87& 41.86& 77.18          & 98.44 &55.55& 67.77          & 91.03 &52.10& 79.81          \\
                                & \textbf{$BW_T$=8, $s_{RM}/s_{CM}$=4}  & 98.74& 48.71& \textbf{65.53} & 92.62 &48.20 &\textbf{79.30}          & 98.24 &63.00 &72.89& 89.65& 61.23& 82.55
\end{tblr}
    \end{table*}

We choose BAF and RCF for comparison, which represent designs with space complexity of O(mn) and O(m+n), respectively. It should be noted that both BAF and RCF are realized with software for sake of fairness because the hardware implementation of same design can vary a lot and our realization of other scheme might be inferior to the original version. Even though, the hardware overhead of different designs can be approximately estimated according to its space complexity.
$N_{CR}$ is set to 1 for all methods, which proves to be effective as shown above. 
In this case, STCF is same with BAF and we therefore leave it. As for SSM, it can be seen as memory optimized version of BAF with downgraded performance but is still O(mn) method essentially, which is also not discussed for simplicity.
Table~\ref{tab:sim_comps} shows the results of different spatiotemporal filter design.
Here, four configurations of CLF are adopted in term of $BW_T$ and $s_{RM}/s_{CM}$.
We can observe that the performance of RCF is inferior to BAF and our design, especially for complex scene. BAF and our design have their own preponderances. For simple scenario like box-moving-2 and box-moving-white, BAF achieves better results. For the other two complex situations, our design achieves better performance of accuracy. The reason we think is that compared with BAF the limited memory capacity of CLF leaves out some noises which avoids mistaking them as events. 
As for different configurations of RCF, generally speaking, more bits the timestamp is represented, and more events the memory unit stores, the better the performance is. However, the performance gap between 8-bit and 32-bit timestamp is trivial for most benchmarks, and sometimes 8-bit timestamp is even better like pendulum. Therefore, reduction the bitwidth of timestamp is a good practice to realize compact implementation on chip.

\subsection{Validation with Recorded DVS Data}
\label{sec:exp_real}
We further validate the performance of the proposed spatiotemporal filter design with recorded DVS data. We employ CeleX-V~\cite{chen2019live} DVS camera for data acquisition. The resolution of sensor is $1280\times 800$, which can be supported by FPGA implementation of RCF. As can be seen from Fig.~\ref{fig:real_data}, four scenarios are selected, including vehicles running on highway (Outdoor-1), pedestrian bridge over road (Outdoor-2), waving hand (Indoor-1), and office environment (Indoor-2). The former three are recorded with stationary camera while the last one moving camera. 40.62M, 4.38M, 42.10M, and 65.23M events are collected, respectively. Then BAF, RCF, and our method are applied for denoising.
The same $T_{th}$ and spatial window of $3\times 3$ are adopted for different methods.
For our design, the configuration of $N_{RM}$/$N_{CM}$=4, $s_{RM}$=4, $s_{CM}$=4, and $BW_{T}$=8 is used.
36.95\%, 76.33\%, 63.46\%, and 55.92\% of the events are removed with our design, respectively.
The visualization of denoising results is also shown in Fig.\ref{fig:real_data}. We can see that the noise of recorded data is effectively removed. However, the quality of visualization image is difficult to judge.
In addition, no ground truth serves as evaluation criterion to quantitatively evaluate the performance of different methods. 
Therefore, we utilize the pre-trained YOLOv5 model to perform car detection on Outdoor-1, which can reflect the effectiveness of denoising in some degree. mAP is calculated as 0.476, 0.453, and 0.486 for BAF, RCF, and our design, respectively. Therefore, we have reason to believe that our RCF design also outperforms previous spatiotemporal filters on real-world DVS data.
\vspace{-3pt}
\section{Conclusion}
\label{sec:conclusion}
This work presents a novel spatiotemporal filter design for DVS denoising, featuring a cache-like memory architecture that exhibits a low space complexity of O(m+n). The pipelining structure of memory access and reduced representation bitwidth of event further reduces power consumption and memory capacity requirement. We implemented our design on FPGA and the experimental results based on both simulated data and recorded data with DVS validate its effectiveness. 
The proposed spatiotemporal filter design provides a low-cost solution for acquiring high-quality DVS event data in real time, thereby facilitating the broader application of DVS technology.
\section{Acknowledgments}
This work was supported by the National Natural Science Foundation of China under Grant 62104182. 

\bibliography{sample-sigconf}

\end{document}